\documentclass[english,10pt,aps,prd,letter,preprintnumbers,floatfix,nofootinbib,showpacs,superscriptaddress]{revtex4}
\usepackage{amsmath, amssymb, latexsym}
\usepackage{hyperref}
\usepackage{slashed}
\usepackage{url}
\usepackage{xcolor}

\catcode`\@=11
\def\lsim{\mathrel{\mathpalette\@versim<}}
\def\gsim{\mathrel{\mathpalette\@versim>}}
\def\@versim#1#2{\vcenter{\offinterlineskip
\ialign{$\m@th#1\hfil##\hfil$\crcr#2\crcr\sim\crcr } }}
\catcode`\@=12

\newcommand{\AddrTUM}{ Physik-Department T30d, Technische Universit\"{a}t M\"{u}nchen.\\
James-Franck-Strasse, 85748 Garching, Germany}
\newcommand{\AddrUNAM}{ Instituto de F\'{\i}sica, Universidad Nacional Aut\'onoma de M\'exico, A.P. 20-364, Ciudad de M\'exico 01000, M\'exico.}
\newcommand{\AddrCol}{Facultad de Ciencias-CUICBAS, Universidad de Colima, C.P.28045, Colima, M\'exico 01000, M\'exico.}
\newcommand{\AddrDual}{Dual CP Institute of High Energy Physics, C.P. 28045, Colima, M\'exico.}

\begin{document}

\title{Dynamical generation of neutrino mass scales}

\author{Alfredo Aranda}\email{fefo@ucol.mx} \affiliation{\AddrCol,\AddrDual}
\author{Cesar Bonilla}\email{cesar.bonilla@tum.de} \affiliation{\AddrTUM}
\author{Eduardo Peinado} \email{epeinado@fisica.unam.mx}\affiliation{\AddrUNAM}

\begin{abstract}

In this letter we present a simple scenario where the mass scales associated to  atmospheric and solar 
neutrino oscillations are obtained through the dynamical generation of neutrino masses. The main idea is that 
the two different scales are the result of two independent mechanisms, namely a type-I seesaw generating the atmospheric 
scale and a radiative 1-loop process providing the solar one. A relation of the two scales, reminiscent of the so-called 
sequential dominance, is thus obtained.

  \end{abstract}


\maketitle


A popular and motivating view among the neutrino physics community is that, since the existence of non-zero neutrino masses is the first clear cut evidence of physics not included in the Standard Model, the smallness of their scale is associated to new physics, commonly dubbed physics beyond the Standard Model. The existence of neutrino masses has been determined through the observation of their oscillations and so far, in the neutrino (lepton) sector, all but one parameter associated to the oscillations have been measured, namely the three mixing angles and the two neutrino mass squared differences (only the absolute value in one of them). The phase associated to CP violation (the one that exists regardless of the fermion nature of neutrinos) is currently being cornered by several experiments and will be determined by future ones~\cite{Abi:2018dnh,Abe:2016ero}. Other very important - and yet unknown - properties of neutrinos that are being explored are their absolute mass scale~\cite{Angrik:2005ep,SejersenRiis:2011sj} (see also ~\cite{Gariazzo:2018pei,deSalas:2018bym})
and fermion nature, i.e. whether neutrinos are Dirac or Majorana fermions, see for 
instance~\cite{Schechter:1980gr,Langacker:2004xy,deGouvea:2004gd,Kayser:2009zz,Deppisch:2012nb}. 

The view that neutrinos might present a window to new physics has motivated a plethora of interesting ideas related to the generation of neutrino masses and their mixing (oscillation) angles. In this letter we present an idea that attempts to relate the two experimentally determined neutrino (squared) mass differences $\Delta m_{atm}$ and $\Delta m_{sol}$. 

There are models in the literature where the two different neutrino squared mass differences are generated with two RH neutrinos in a sequential dominance way, see for instance~\cite{King:1999cm,Antusch:2004gf,King:2015dvf}. Our approach is similar in spirit and attempts to find a relation among the two scales that might lead to interesting new physics scenarios, in particular with regards to dark matter. The basic idea consists of having two independent mechanisms for the generation of neutrino mass that lead to the observed ratio of scales. The atmospheric scale turns out to come from a type-I seesaw while the solar scale is a product of a radiative 1-loop generation of neutrino mass.

We now proceed to describe the specific scenario: In addition to the SM field content and symmetries, we incorporate two right-handed (RH)
neutrinos $N$ and $N^\prime$, one scalar $SU(2)_L$ singlet charged under lepton number $\phi$, and one extra ``Higgs" $SU(2)_L$ doublet $\eta$. The scalar 
field $\phi$ acquires a vacuum expectation value (vev) $v_{\phi}$ breaking lepton number and dynamically giving a mass to the RH neutrinos. An additional
$\mathbb{Z}_2$ symmetry is imposed under which only $\eta$ and $N^\prime$ are charged (odd), thus making them the ``dark sector" of the model.
The relevant quantum numbers for the fields participating in the generation of neutrino masses are listed in Table~\ref{tab:Model}.
\begin{table}[!h]
\setlength\tabcolsep{0.25cm}
\centering
\begin{tabular}{| c | c | c | c |c|}
\hline
                      &  $N$   & $\eta$   & $N^\prime$ & $\phi$ \\   
\hline
\hline                                    
$\mathrm{SU(2)_L}$    &   1    &    2    &    1   & 1  \\
$\mathrm{U(1)_Y}$     &   0    &   1/2   &    0   & 0  \\
$U(1)_L$          &   1    &    0    &   1    & -2  \\
$\mathbb{Z}_2$        &  $+$   &   $-$   &  $-$  &  0 \\
\hline
\end{tabular}
\caption{\label{tab:Model} Field content and transformation properties (charges) of the model for fields additional to the SM ones. All SM fields are $\mathbb{Z}_2$ even.}
\end{table}

 The Lagrangian for this model is
$\mathcal{L} = \mathcal{L_{\rm SM}} + \mathcal{L}_{kin}(N,N',\eta,\phi) +\mathcal{L_{\rm ATM}}  + \mathcal{L_{\rm DM,SOL}} - V(H,\eta,\phi) $, where $\mathcal{L_{\rm SM}}$ is the Standard Model Lagrangian, $\mathcal{L}_{kin}(N,N',\eta,\phi)$ contains the kinetic terms of the new fields, $ \mathcal{L_{\rm ATM}}$ is given by
\begin{eqnarray}
\mathcal{L_{\rm ATM}} &=& - Y^{(0)}_i \overline{L}_i \tilde{H} N\,  +\, Y^{N} \phi \overline{N^c}N\,  +~h.c. \label{eq:atm}
\end{eqnarray}
with $\tilde{H} = i \tau_2 H^*$, $i={1,2,3}$, and 
\begin{eqnarray}
  \mathcal{L_{\rm DM,SOL}} &=& Y^{(1)}_i \overline{L}_i \tilde{\eta} N^\prime\, + Y^{N'} \phi \overline{N^{\prime c}}N^\prime +~h.c. \label{eq:solar}
\end{eqnarray}
with $\tilde{\eta} = i \tau_2 \eta^*$ and $i={1,2,3}$. As we will see below, the Lagrangian $\mathcal{L_{\rm ATM}} $ in eq. (\ref{eq:atm}) induces an effective non-zero tree level neutrino mass once the electroweak symmetry is broken by the vacuum expectation value of the Standard Model Higgs. This scale is identified with the atmospheric neutrino scale. On the other hand, the Lagrangian $\mathcal{L_{\rm DM,SOL}}$ in eq. (\ref{eq:atm}) is responsible for the solar neutrino scale  {\it \`a la  scotogenic}, namely through a 1-loop process involving the scalar $\eta$~\cite{Ma:2006km}.

Finally, the scalar potential $V(H,\eta,\phi)$ is given by
\begin{align}
V(H,\eta,\phi) &=  \mu_1^2 H^\dagger H + \mu_2^2 \eta^\dagger \eta + \mu_3^2 \phi^* \phi + \lambda_1 (H^\dagger H)^2 + \lambda_2 (\eta^\dagger \eta)^2 + \lambda_3 (\eta^\dagger \eta)(H^\dagger H) \nonumber
+ \lambda_4 (\eta^\dagger H) (H^\dagger \eta ) \\ &+ \frac{\lambda_5}{2} \left( (\eta^\dagger H)^2 + (H^\dagger \eta)^2 \right) + \lambda_6 (\phi^* \phi)^2 
+ \lambda_7 (\phi^* \phi) (H^\dagger H) + \lambda_8 (\phi^* \phi) (\eta^\dagger \eta).
\label{Eq.ScalarPot}
\end{align}

The spontaneous breaking of both lepton number and electroweak symmetries is triggered by the vev's of $H$ and $\phi$ respectively and lead to the following scalar mass spectrum: two CP-even fields coming from $H$ and $\phi$ with masses
\begin{equation}
M^2_{(h_1,h_2)} = \left(v_H^2 \lambda_1 + v_\phi^2 \lambda_6\right) \mp \sqrt{v_H^2v_\phi^2\lambda^2_7 +(v_H^2\lambda_1 - v_\phi^2\lambda_6)^2},
\end{equation}
with the ``$-$" (``$+$") in $\mp$ corresponding to $h_1$ ($h_2$), 
a physical Nambu-Goldstone boson resulting from the breaking of the $U(1)_L$ symmetry, the Majoron $J=Im(\phi)$; a third CP-even scalar as well as its CP-odd {\it companion} coming from the {\it inert} doublet $\eta$ with masses
\begin{equation}
M^2_{(\eta_R,\eta_A)}=\frac{1}{2}\left(\mu^2_2+ \lambda_8 v_\phi^2 + (\lambda_3+\lambda_4\pm \lambda_5)v_H^2\right)
\end{equation}
with the ``$+$" (``$-$") in $\pm$ corresponding to $\eta_R$ ($\eta_A$). Note that $\lambda_5 v_H^2 = (M^2_{\eta_R} - M^2_{\eta_A})$; and finally a charged scalar field with mass
\begin{equation}
M^2_{\eta^\pm}=\frac{1}{2}\left(\mu_2^2 + \lambda_3 v_H^2 + \lambda_8 v_\phi^2\right) .
\end{equation}
 In the fermion sector, at tree level, the contribution from type I see-saw leads to the following {\it leading} neutrino mass matrix 
\begin{equation}
{\cal M}_\nu^{(0)} = -v_H^2 Y^{(0)}M_{N}^{-1} (Y^{(0)})^{T} 
\label{eq:TI}
\end{equation}
where $M_N=Y^N v_\phi$. Setting the vector of Yukawa couplings as  $Y^{(0)}=(y_1,y_2,y_3)$, eq. (\ref{eq:TI}) takes the matrix form
\begin{equation} 
\label{eq:MnuA}
{\cal M}_\nu^{(0)} 
 =-\frac{v_H^2}{Y^N v_\phi} 
\begin{small}
 \begin{pmatrix}
 \left(Y^{(0)}_1\right)^2 & Y^{(0)}_1 Y^{(0)}_2  &  Y^{(0)}_1 Y^{(0)}_3  \\
 Y^{(0)}_1 Y^{(0)}_2 & \left(Y^{(0)}_2\right)^2  &  Y^{(0)}_2 Y^{(0)}_3  \\
 Y^{(0)}_1 Y^{(0)}_3 & Y^{(0)}_2 Y^{(0)}_3  &  \left(Y^{(0)}_3\right)^2  
  \end{pmatrix}.
\end{small}
\end{equation}

This is a rank-1 matrix and therefore gives a non-zero eigenvalue to be associated to the {\it heaviest} neutrino (we are assuming normal ordering where $m_3 > m_2 > m_1$).

At the one-loop level, $\eta$ allows the radiative generation of neutrino masses with a contribution given by~\cite{Ma:2006km}:
\begin{equation}
({\cal M}^{(1)}_\nu)_{ij} = \frac{Y^{(1)}_{i}Y^{(1)}_{j} M_{N^\prime}}{16 \pi^2}\left[ \frac{M_{\eta_R}^2}{M_{\eta_R}^2-M_{N^\prime}^2}\text{ln}\frac{M_{\eta_R}^2}{M_{N^\prime}^2} -  \frac{M_{\eta_A}^2}{M_{\eta_A}^2-M_{N^\prime}^2}\text{ln}\frac{M_{\eta_A}^2}{M_{N^\prime}^2}           \right] \ ,
\label{mnuloop}
\end{equation}
where $M_N^{\prime}=Y^{N^\prime} v_\phi$. 
Assuming that the mass difference between $\eta_R$ and $\eta_I$ is small compared to $M_0^2=(M_{\eta_R}^2+M_{\eta_A}^2)/2$, one 
gets
\begin{equation}
({\cal M}^{(1)}_\nu)_{ij} = \frac{\lambda_5 v_H^2}{16 \pi^2} Y^{(1)}_{i}Y^{(1)}_{j} \,f(M_0,M_{N^{\prime}}) 
\ \ \text{where} \ \ f(M_0,M_{N^{\prime}})=\frac{M_{N^{\prime}}}{\left(M_0^2-M_{N^\prime}^2\right)}
\left[ 1 - \frac{M_{N^\prime}^2}{\left(M_0^2-M_{N^\prime}^2\right)}\text{ln}\frac{M_0^2}{M_{N^\prime}^2}\right].
\label{mnuloopm0}
\end{equation}

The matrix in eq.~(\ref{mnuloop}) turns out to be a rank-1 matrix giving another non-zero 
eigenvalue (to be associated with $m_2$). Note that our set up is based on the fact that the Yukawa vectors lead to rank-1 
neutrino mass matrices and thus the lightest neutrino state is massless. In order to avoid this situation in a complete model one would need to introduce more fields in order to generate a very small mass for it. 

In order to see the relation between the two scales in neutrino oscillation parameters, we define
\begin{equation}
\label{rnu}
R_{\nu}\equiv \left[ {\Delta m^2_{\text{atm}} \over \Delta m^2_{\text{sol}}} \right]^{1/2}
\sim
{\Delta m_{\text{atm}} \over \Delta m_{\text{sol}}}.
\end{equation}

Using the fact that
\begin{equation}
\label{datmsol}
\Delta m_{\text{atm}} \sim \left({ \frac{v_H^2 \hat{Y} }{M_{N}}}\right) \ \ \text{and}
\ \
\Delta m_{\text{sol}} \sim \left(\frac{\tilde{Y} \lambda_5 v_H^2 }{16 \pi^2}\right) \times f(M_0,M_{N^{\prime}}),
\end{equation}
where $\hat{Y}=\sum \left(Y^{(0)}_i\right)^2$ and $\tilde{Y}=\sum \left(Y^{(1)}_i\right)^2$. 

One can write $M_0=\alpha M_{N^{\prime}}$ which leads $f(M_0,M_{N^{\prime}})$ to
\begin{equation}
f'=\frac{1}{M_{N'}}\frac{1}{\left(\alpha^2-1\right)}
\left[ 1 - \frac{1}{\left(\alpha^2-1\right)}\text{ln}\,\alpha^2 \right].
\end{equation}
Then, using the relations in eq.~(\ref{datmsol}),
the ratio in eq.(\ref{rnu}) becomes  
\begin{equation}
R_{\nu} \sim \left( {16 \pi^2 \over \lambda_5} {\hat{Y} Y_{N'} \over \tilde{Y} Y_{N}} \right)\times \text{number}.
\end{equation}

From this is clear that if we take $\lambda_5\sim\mathcal{O}(1)$
then the product $\hat{Y}Y_{N'}$ must be smaller than the product $\tilde{Y}Y_{N}$ in order to explain 
the experimental data (note that the hierarchy we have in the Yukawas would correspond to the so called 
sequential dominance~\cite{Antusch:2004gf,King:2015dvf} of the RH neutrino masses).  We would like to mention
that while preparing this work, a paper appeared in the arXiv where a very similar approach was suggested for 
a relation between the two scales~\cite{Rojas:2018wym}. The main difference in our set up is the dynamical
generation of the mechanism, something we consider important for it gives a glimpse on the possible origin of
the assumptions needed to generate the rank-1 matrices.
 
We conclude this letter with some comments regarding dark matter. In this set up the dark matter candidate turns out to be the lightest $\mathbb{Z}_2$-odd particle, namely,  either the scalar ($\eta_R,\eta_A$) or the 
Majorana fermion $N'$. Note that if it is the scalar the situation is similar to the Inert Doublet Model~\cite{Belyaev:2016lok} and the
dark matter constrains are the same as in that case. If the RH neutrino $N'$ is the candidate, and taking as an 
example $m_0^2\simeq 2 M_{N'}^2$, eq.~(\ref{rnu}) reduces to
\begin{equation}
R_{\nu} \sim \left( {16 \pi^2 \over \lambda_5} {\hat{Y} Y_{N'} \over \tilde{Y} Y_{N}} \right)\times3.
\end{equation}

In order to be in agreement with the observed hierarchy $R_\nu\sim \sqrt{30}$, one can take for 
instance $\lambda_5=[0.1,1]$ and hence the ratio between the Yukawas turns out to be
$r={\hat{Y} Y_{N'} \over \tilde{Y} Y_{N}}\sim [10^{-3},10^{-2}]$. These ratios are easily obtained 
when the two RH neutrinos are close in mass, i.e. $Y_{N'}\sim Y_{N}$, and there is a hierarchy between 
Yukawa couplings $Y^{(0)}_i$ and $Y^{(1)}_i$ around one order of magnitude, namely $Y^{(0)}_{i}\sim 0.1 {Y^{(1)}_i}$.
Here the smallness of neutrino masses could in principle be explained from the heaviness of the neutral scalars.
 It is worth to mention that since the charged lepton mass matrix is a general complex, there is enough freedom to accomodate  neutrino oscillation data~\cite{Esteban:2016qun,deSalas:2017kay}.
In the {\it generic} fermionic DM case there is strong  fine-tuning in the Yukawa couplings~\cite{Ibarra:2016dlb}
due to the fact that, on the one hand, since DM annihilation into leptons is through the t-channel, and in order to have a correct relic density, those Yukawas cannot be not very small, yet, on the other hand,  the same couplings  generate neutrino 
masses and lead to lepton flavor violating processes such as $\mu\rightarrow e+\gamma$ and $\mu\rightarrow 3e$  (for example) 
and thus must be small~\cite{Toma:2013zsa,Vicente:2014wga,Ibarra:2016dlb}.

In our set up, however,  there is an additional DM annihilation channel (namely, the s-channel) due to the presence 
of the scalar singlet~\cite{Babu:2007sm,Bonillaetal} responsible of breaking lepton number, making the possibility of  a fermion DM candidate
very likely, in contrast to what happens in the minimal scotogenic model. In fact, in this case there exist an  
annihilation channel of the DM candidate into Majorons that can be controlled to guarantee detectability. The detailed analysis of such a scenario  will be presented in~\cite{Bonillaetal}.

\begin{acknowledgments}
The work of A.A. was supported by CONACYT project CB-2015-01/257655 (M\'exico). CB has  
been supported by the Collaborative Research Center SFB1258. EP acknowledges financial support from  DGAPA-PAPIIT IN107118, the German-Mexican research collaboration grant SP 778/4-1 (DFG) and 278017 (CONACyT) and PIIF UNAM.
\end{acknowledgments}

\end{document}